\documentclass[useAMS,amssymb,epsfig,usegraphicx,usenatbib,twocolumn,revtex4-1]{emulateapj}

\def\aj{AJ}

\def\apj{ApJ}
\def\apjl{ApJ}
\def\apjs{ApJS}
\def\aap{A\&A}
\def\mnras{MNRAS}
\def\na{NewA}

\usepackage[usenames,dvips]{color}

\newif\ifAMStwofonts

\makeatother

\shorttitle{Why are some galaxies not barred?}
\shortauthors{Saha \& Elmegreen}

\begin{document}
\title{Why are some galaxies not barred?}

\author {Kanak Saha$^{1}$ \& Bruce Elmegreen$^{2}$}
\affil{$^{1}$ Inter-University Centre for Astronomy and Astrophysics, Pune 411007, India, \\
$^{2}$ IBM Research Division, T. J. Watson Research Center, 1101 Kitchawan Road, Yorktown
Heights, NY 10598, USA
\\e-mail: kanak@iucaa.in, bge@us.ibm.com}

\label{firstpage}

\begin{abstract}
Although more than two-thirds of star-forming disk galaxies in the local universe are
barred, some galaxies remain un-barred, occupying the upper half of the Hubble tuning
fork diagram. Numerical simulations almost always produce bars spontaneously, so it
remains a challenge to understand how galaxies sometimes prevent bars from forming. Using 
a set of collisionless simulations, we first reproduce the common result that
cold stellar disks surrounding a classical bulge become strongly unstable to
non-axisymmetric perturbations, leading to the rapid formation of spiral structure and
bars. However, our analyses show that galaxy models with compact classical bulges (whose 
average density is greater than or comparable to the disk density calculated within 
bulge half-mass radii) are able to prevent bar formation for at least 4 Gyr even when the 
stellar disk is maximal and having low Toomre Q. Such bar prevention is the result of 
several factors such as (a) a small inner Lindblad resonance with a high angular rate, 
which contaminates an incipient bar with $x_2$ orbits, (b) rapid loss of angular momentum 
accompanied by a rapid heating in the center from initially strong bar and spiral instabilities
 in a low-Q disk, in other words, a rapid initial rise to a value larger than $\sim5$ of the ratio 
of the random energy to the rotational energy in the central region of the galaxy.
\end{abstract}


\section{Introduction}
\label{sec:intro} 

\noindent Stellar bars are one of the most common non-axisymmetric structures
in spiral galaxies. More than $60\%$ of disk galaxies in the local universe are
strongly barred \citep{Eskridgeetal2000,Grosboletal2004, MenendezDelmestreetal2007,
Barazzaetal2008}. Our local group is no exception to this. Bars are also seen out to
redshifts $z \sim 1$ \citep{sheth08}, corresponding to 8 Gyr ago, which implies that
once formed, a bar is hard to destroy \citep{Athanassoulaetal2005}. Bars also seem to
have formed relatively quickly, as they appeared soon after galaxy disks became cool
\citep{sheth12}. Thus it remains unclear why all galaxies are not barred.

\noindent Most of our knowledge about bar formation has come from numerical simulations, starting
with \cite{miller70} and \cite{hohl71}. Simulations of isolated galaxies with cool
stellar disks show the spontaneous formation of bars from gravitationally unstable
$m=2$ modes \citep{Toomre1964, Goldreich-Tremaine1979, Toomre1981, CombesSanders1981,
SellwoodWilkinson1993, Polyachenko2013}, or from galaxy interactions and mergers
\citep{Noguchi1987,Gerin1990,Elmegreen1991,BarnesHerquist1991,MiwaNoguchi1998} or
interactions with dark matter halo substructures \citep{Romano-Diazetal2008}. Resonant
gravitational interactions that transfer disk angular momentum to the dark matter halo
lead to long-term bar stability and growth \citep{DebattistaSellwood1998,
Athanamisi2002, Athanassoula2003, Holley-Bockelmannetal2005, WeinbergKatz2007a,
Ceverinoklypin2007, Dubinskietal2009, Sahaetal2012}. Even hot stellar disks that are
otherwise stable can form bars if they are embedded in a spinning dark matter halo
\citep{SahaNaab2013}.

There have been a number of studies that addressed this issue by finding processes that
can destroy a bar. The usual suspects are central mass concentrations (CMC) and
super-massive black holes (SMBHs), possible fed by gas inflow
\citep{BournaudCombes2002}. Both CMCs and SMBHs can affect the orbital distribution of
stars in a bar and dissolve the bar on timescales of a few Gyr or less
\citep{PfennigerNorman1990,Hasanetal1993,HozumiHernquist2005}. However, other studies
suggest that bars are difficult to destroy because the central mass has to be
unreasonably large, such as $\sim10\%$ of the disk mass \citep{ShenSellwood2004,
Athanassoulaetal2005}. Also, CMCs and SMBHs are present in barred galaxies, so they
seem to co-exist. What has yet to be determined is whether a bar can grow in the first
place in the presence of such compact objects. Perhaps the prevention of bar formation
at an early stage is easier than the destruction of the bar after it gets massive.

The present paper uses self-consistent simulations that probe the impact of compact
classical bulges on bar growth. Sec~\ref{sec:modelsetup} describes the simulated
galaxies and Sec~\ref{sec:barform} discusses bar formation and various early effects
introduced by a bulge. Sec~\ref{sec:dissection} investigates in detail, the dynamics and 
evolution of two models. The discussion and conclusions are in Sec~\ref{sec:discussion}.

\begin{figure}
\rotatebox{0}{\includegraphics[height=13. cm]{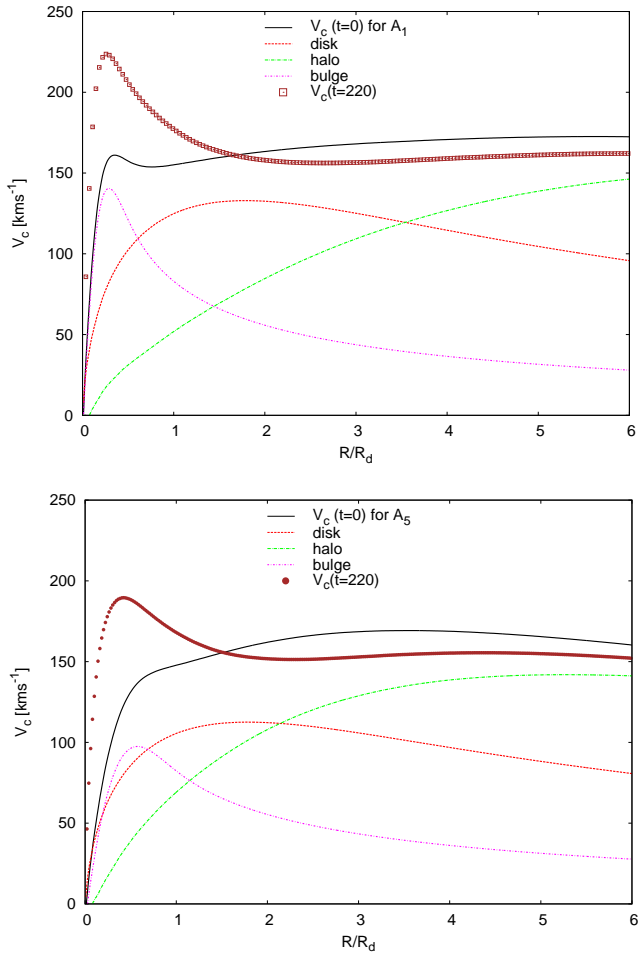}}
\caption{Circular Velocity curves for model $A_1$ (upper) and $A_5$ (lower). Time is in internal unit.}
\label{fig:Vcirc}
\end{figure}

\begin{figure*}
\begin{flushleft}
\rotatebox{0}{\includegraphics[height=7.0 cm]{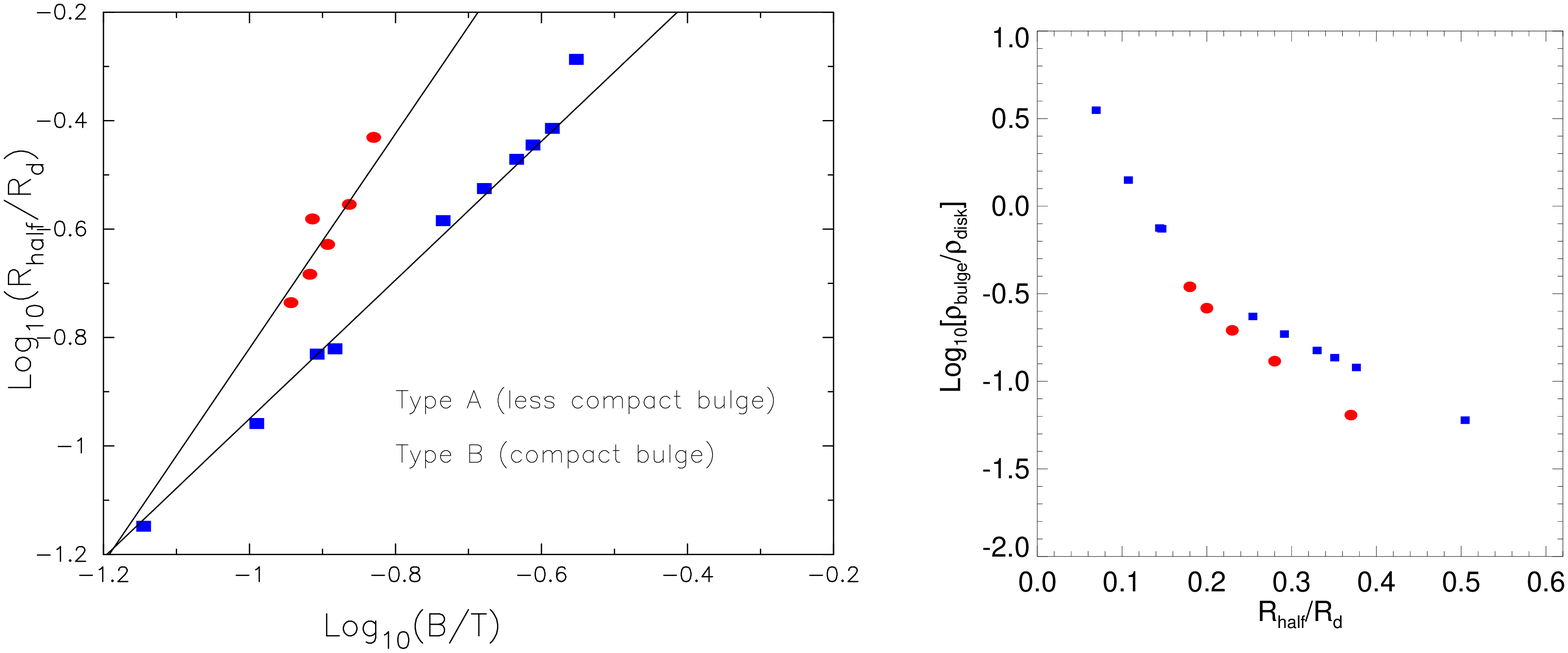}}
\caption{Left: Mass-size relation for two different sets of classical bulges, type~A
and type~B. For a given $B/T$, type~B bulges are more compact than type~A. 
Right: Bulge density normalized by the disk density within $R_{half}$ plotted against 
their half-mass radii. Two bulges from type~B have densities higher than their respective 
disk density (see table~\ref{tab:paratab}).} 
\label{fig:mass-radius}
\end{flushleft}
\end{figure*}

\section{galaxy model and simulation}
\label{sec:modelsetup}

We construct a set of $15$ three-component galaxy models consisting of
a stellar disk, a dark matter halo, and a classical bulge, initially in equilibrium.
The stellar disk is initially axisymmetric with surface density following an
exponential profile with scale length $R_d$ and central surface density $\Sigma_0$. The
initial vertical scale height is kept constant for simplicity and follows a $sech^{2}$~distribution,
with scale-height $h_z=0.02 R_d$ for all models. For self-consistency, we let the
vertical velocity dispersion follow also an exponential profile with a scale length
equal to twice that of the surface density. The dark matter halo is modelled with a
lowered \cite{Evans93} model, which produces a nearly flat circular velocity profile
, see Fig.~\ref{fig:Vcirc} for two models, $A_1$ and $A_5$. The bulge is modelled 
with a King distribution function \citep[DF,][]{King1966} and their properties are discussed below.
Further details about the distribution function and model construction can be found 
in \cite{KD1995} and \cite{Sahaetal2012}. The dark matter halo mass is kept nearly 
constant in most models at about $M_h \simeq 4 M_d$; slight variations are reported 
in Table~\ref{tab:paratab}. The stellar disk is the same in all models. The mass and size 
of the initial classical bulge vary the most from model to model. 

\begin{table}[b]
\caption[ ]{Initial parameters of the galaxy models.}
\begin{flushleft}
\begin{tabular}{ccccccc}  \hline\hline
Models     & $M_h/M_d$ & $B/T $ & $R_{half}/R_d$ & $Q_{2.5}$ & $\rho_{bulge}$ & $\rho_{disk}$ \\
       & &  &  & & &\\

\hline
$A_1$   & 4.3 & 0.114 & 0.18 & 1.208 & 4.88 & 14.1\\
$A_2$   & 4.2 & 0.121 & 0.20 & 1.280 & 3.63 & 13.9\\
$A_3$   & 4.2 & 0.128 & 0.23 & 1.300 & 2.65 & 13.6\\
$A_4$   & 4.1 & 0.137 & 0.28 & 1.335 & 1.72 & 13.2\\
$A_5$   & 4.0 & 0.148 & 0.37 & 1.437 & 0.80 & 12.5\\
\hline
\hline
$B_1$ & 4.5 & 0.07  & 0.07 & 0.953 & 53.7 & 15.2\\
$B_2$ & 4.2 & 0.10  & 0.10 & 1.043 & 21.0 & 14.9\\
$B_3$ & 4.3 & 0.121  & 0.144 &1.099 & 10.8 & 14.4\\
$B_4$  & 4.2 & 0.128  & 0.147 &1.107 & 10.7 & 14.4\\
$B_5$ & 4.0 & 0.18  & 0.25 & 1.097 & 3.2 & 13.5\\
$B_6$ & 4.0 & 0.20  & 0.29 &1.136 & 2.4 & 13.1\\
$B_7$ & 3.5 & 0.23  & 0.33 &1.185 & 1.92 & 12.8 \\
$B_8$ & 4.2 & 0.24  & 0.35 &1.154 & 1.72 & 12.6\\
$B_9$ & 3.5 & 0.25  & 0.37 &1.164 & 1.50 & 12.5\\
$B_{10}$ & 4.1 & 0.27  & 0.50 &1.198 & 0.69 & 11.5\\
\hline
\end{tabular}
\end{flushleft}
\label{tab:paratab}
\end{table}

\begin{figure}
\begin{flushleft}
\rotatebox{270}{\includegraphics[height=8.5 cm]{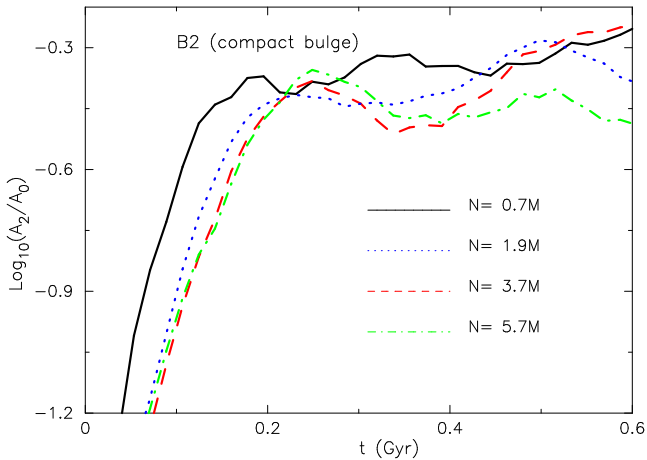}}
\caption{Time evolution of the $A_2/A_0$ for model $B_2$ and dependence on the
number of particles used in the simulation.}
\label{fig:convergence}
\end{flushleft}
\end{figure}

\subsection{Properties of model classical bulges}
\label{sec:ClBs}

A spherical live classical bulge (hereafter ClBs) is constructed from the King model. 
For the sake of completeness, the corresponding DF is given by 
\citep{KD1995}

\begin{equation}
  f_{b}(E)=\left\{
    \begin{array}{ll}
      \rho_{b}(2\pi\sigma_{b}^2)^{-3/2}
      e^{(\Psi_{b0}-\Psi_{c})/\sigma_{b}^2} &\\
      \times \{e^{-(E-\Psi_{c})/\sigma_{b}^2}-1\}
        & \mbox{if} \; E < \Psi_{c},\\
      0  \quad &\mbox{otherwise}.
    \end{array} \right.
\end{equation}

\noindent Here, the bulge is specified by three parameters, namely the cut-off 
potential ($\Psi_c$) which determines the bulge tidal radius, central bulge density 
($\rho_b$) and central bulge velocity dispersion ($\sigma_b$). The gravitational 
potential at the centre of the bulge is measured by $\Psi_{b0}$. The radial density 
profile has a core at the centre and sharply drops to zero at the tidal radius. 
The more negative the $\Psi_c$, the more centrally concentrated and radially confined is 
the bulge. The parameters $\sigma_{b}$ and $\rho_{b}$ control the mass and size 
of the bulge. Initially, we picked a range of values for these parameters, namely 
$\rho_{b} ={3.4 - 250.8}$, $\sigma_{b}={0.7 - 2.5}$. The value of $\Psi_c$ was 
varied the least, from $-2.8$ to $-3.2$. 
Most of B-series models (see below) had higher $\rho_{b}$ and $\Psi_{c}$, e.g., for
the $B_1$ model they are $250.8$ and $-3.2$ respectively. 
Note that there is no one-to-one correspondence
between these parameters of the DF and the mass model of the bulge (e.g., bulge mass and 
tidal radius), since the bulge is gravitationally coupled with the other two components, 
disk and halo. Not all sets of parameters lead to convergence when creating a galaxy model, 
see \citep{KD1995} and its user manual for GalactICs.

In the left panel of Fig.~\ref{fig:mass-radius}, we show the relation between the 
bulge size (measured by $R_{half}$, the bulge half-mass radii) and the bulge-to-total ($B/T$) 
ratio for each galaxy model, where $B$ denotes the bulge mass and $T$ denotes the total 
stellar mass for our simulated galaxy models. To these simulated ClBs, we fit the following 
linear regression model between $log[R_{half}]$ and $log[B/T]$:

\begin{equation}
R_{half}/R_{d} = C_{0} (B/T)^{\alpha},
\end{equation}

\noindent  where $C_0$ is a constant and $\alpha$ is the power-law exponent.
Fig.~\ref{fig:mass-radius} shows two tracks for the distribution of the bulge size
with the B/T ratio. We call these type A and B models, with type A representing the 
less compact bulges and type B representing the more compact bulges.  

\begin{figure*}
\rotatebox{0}{\includegraphics[height=11.0 cm]{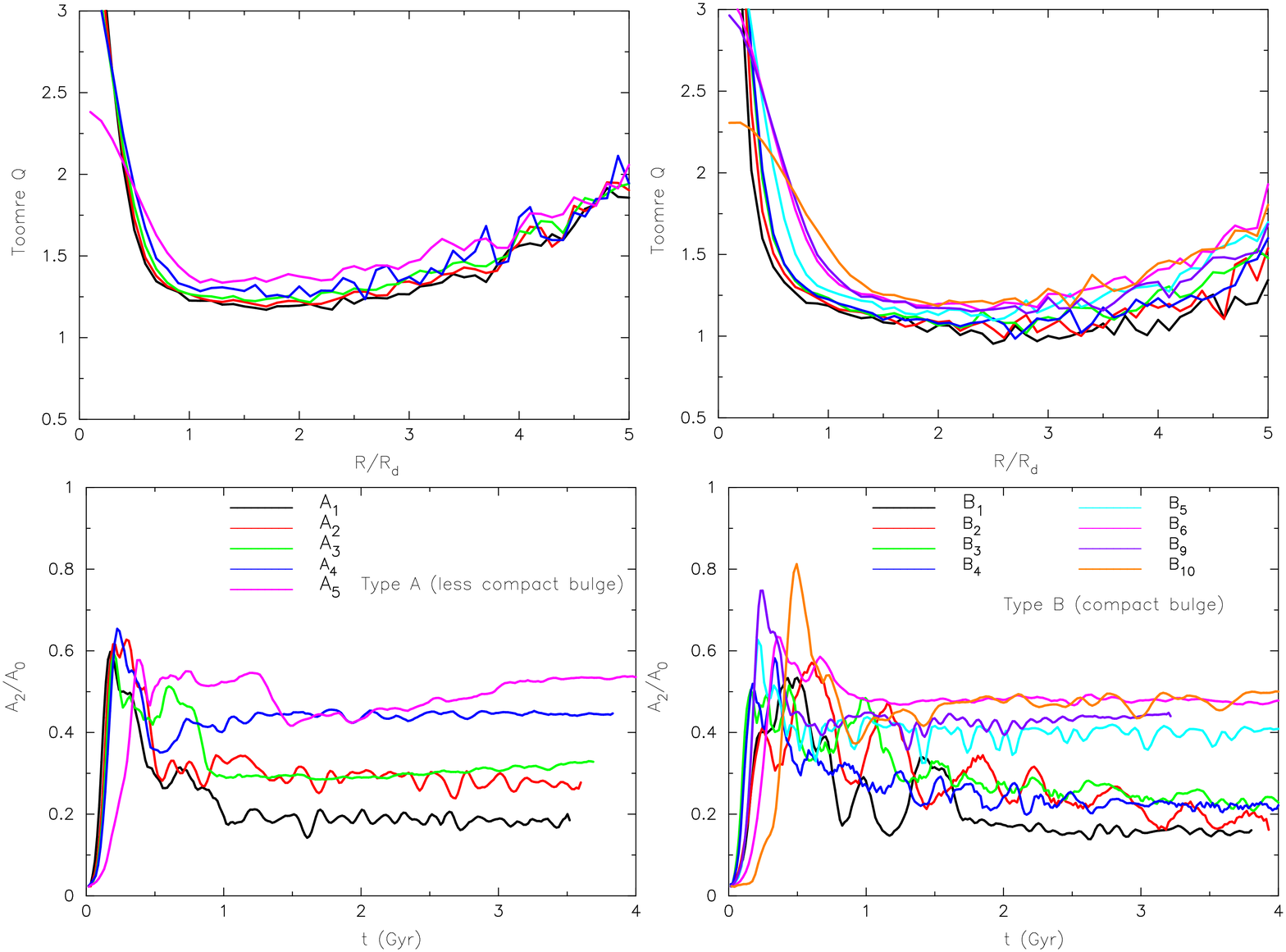}}
\caption{Left panels: initial radial profile of Toomre $Q$ (upper) and time 
evolution of $A_2/A_0$ (lower) for type~A models. Right panels: same as left panels but for
type~B.}
\label{fig:QA2vstime}
\end{figure*}

\begin{figure*}
\rotatebox{0}{\includegraphics[height=11.0 cm]{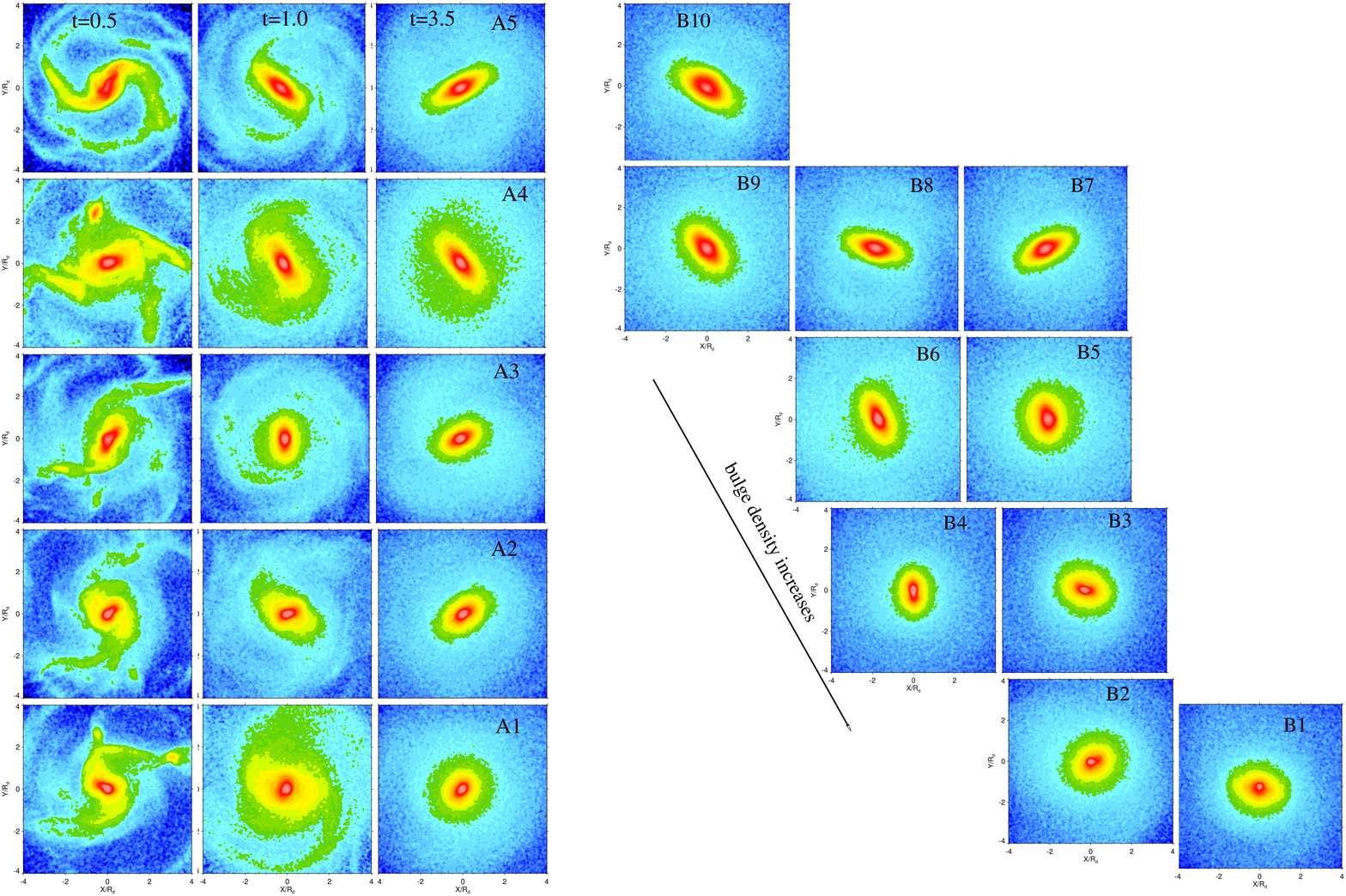}}
\caption{Left panels: face-on surface density maps and their time sequence for type-A models.
Model with highest density ClB in A-series (e.g., $A_1$) has no bar at the end of $3.5$~Gyr. 
From upper panels to lower ones, the bulge density increases, making it difficult to grow bar 
at the highest density. Right panels: Same but for type~B model galaxies at $t=3.5$~Gyr. Models are
arranged according to their bulge density.}
\label{fig:faceon}
\end{figure*}

For type~A ClBs, we have $\alpha=1.98$ and $C_0 = 1.16$. For type~B ClBs, 
they are $1.28$ and $0.33$ respectively. A similar relation holds for the outer 
radii ($R_b$) of the King bulges and $B/T$, with $\alpha = 2.0$, $C_0=1.73$ for 
type~A ClBs. For type~B, they are $1.2$ and $0.85$ respectively. 
So for a given $B/T$, type~B bulges are more compact than type~A bulges. 
On the right panel of Fig.~\ref{fig:mass-radius}, we show the dependence of the average bulge density
on the half-mass radius. The bulge density is computed as
$\rho_{bulge}= {3 M_{b,{1/2}}}/{(4\pi R_{half}^3)}$, where $M_{b,{1/2}}$ is the ClB mass 
within $R_{half}$. These quantities are calculated exclusively using the
bulge particles (following their unique id's in the simulation).
We normalize the bulge density with the density of disk stars ($\rho_{disk}$) measured 
within $R_{half}$, see Table~\ref{tab:paratab}. This ratio is given by:

\begin{equation}
\frac{\rho_{bulge}}{\rho_{disk}} = \frac{M_{b,{1/2}}}{M_d} \frac{3 h_z}{4 R_{half}}
[1- e^{-R_{half}/R_d}\times(1 + R_{half}/R_d)]^{-1},
\end{equation}

The right panel of Fig.~\ref{fig:mass-radius} demonstrates that type~A bulges are less dense 
compared to type~B. In other words, we can say that type~B bulges are more compact and 
dense than type~A.

We scale the models such that $R_d = 3$~kpc and the circular velocity at $2 R_d$ is
$160$~km/s. The unit of time varies from model to model as the bulge mass varies. 
We have used a total of $3.7 \times 10^{6}$ particles, with $2.0 \times 10^{6}$ for the dark matter halo,
$1.2 \times 10^{6}$ for the disk and $0.5 \times 10^{6}$ for the bulge particles.
The softening lengths for the disk, bulge and halo particles are calculated following the suggestion 
of \cite{McMillan2007}. The simulation is performed using the Gadget-1 code \citep{Springeletal2001}, 
which uses the quadrupole contribution to the force calculation, using a
tolerance parameter $\theta_{tol} =0.7$ and an integration time step of $0.03$ times
the internal time unit. The simulation was evolved for a time period of $\sim 4.0$ Gyr. The energy 
is conserved within $0.1\%$ and angular momentum within $1 \%$ for entire
duration of the run. 

We have also run a few more simulations to check convergence with 
respect to the number of particles. In particular, we have re-run model $B_2$ which evolved into 
an unbarred galaxy, with particle numbers varying from 0.7 million to 5.7 million. We noticed that
increasing particle number delays the linear growth as expected; we found $\sim 50$~Myr of delay 
as the particle number was increased by a factor of 8. Fig.~\ref{fig:convergence} demonstrates that 
the linear growth phase remains nearly same between $N=3.7$ and $5.7$ million particles. 
The convergence on the number of particles in our simulation is in compliance with previous
results found in \cite{Dubinskietal2009} and \cite{Sahaetal2010}. We also noticed that model
$B_2$ evolved always into an unbarred case irrespective of the particle number, provided it was about
a few million.   

\begin{figure}[b]
\begin{flushleft}
\rotatebox{270}{\includegraphics[height=8.0 cm]{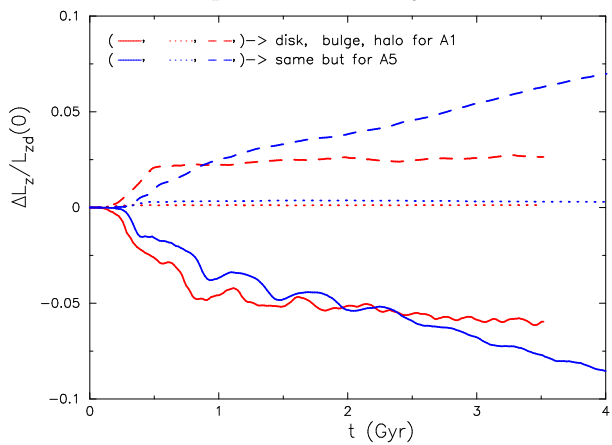}}
\caption{Angular momentum exchange between the stellar disk and dark matter halo for models
$A_1$ (red curves) and $A_5$ (blue curves). $L_{zd}(0)$ is the z-component of the disk angular
momentum at t=0. The gain of angular momentum by the dark halo of $A_1$ saturates after 0.5~Gyr.}
\label{fig:AM}
\end{flushleft}
\end{figure}

\begin{figure*}
\rotatebox{0}{\includegraphics[height=11. cm]{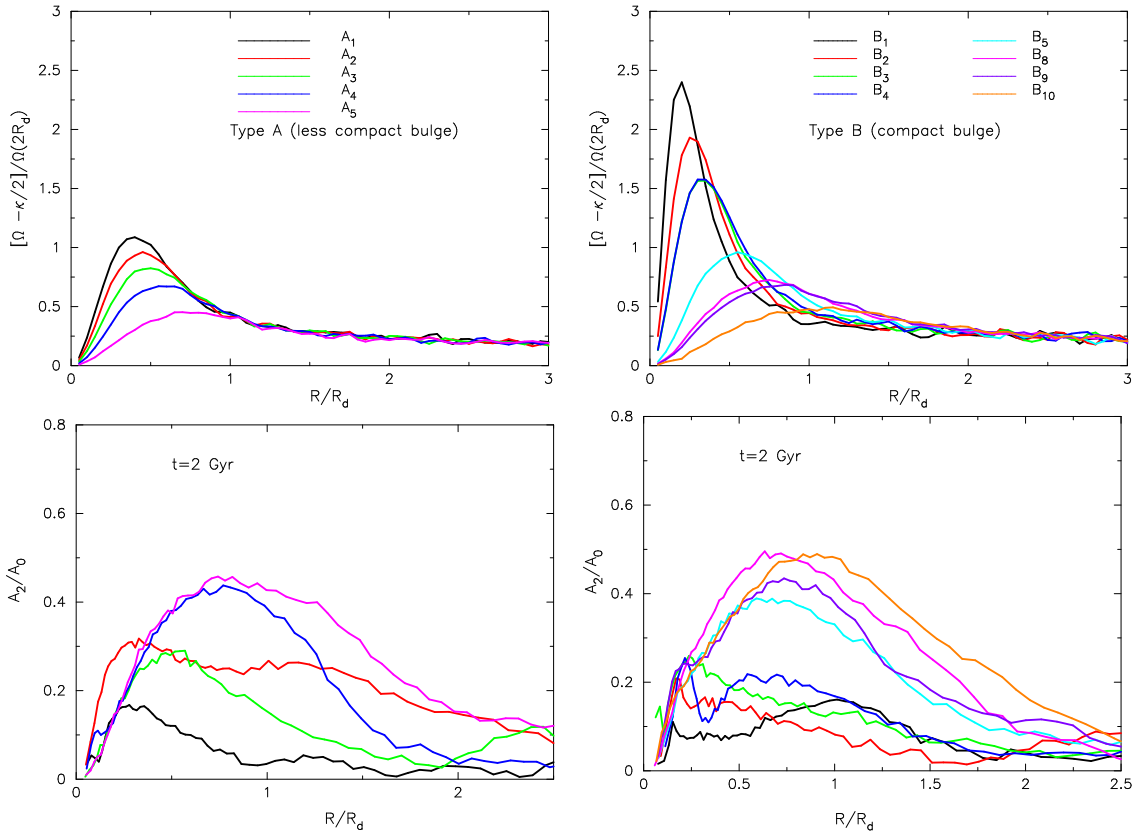}} 
\caption{Left panels: initial ILR curves (upper) and radial profile of $A_2/A_0$ (lower) 
for type~A models. Right panels: same as the left panels but for type~B} 
\label{fig:ILRA2vsR}
\end{figure*}

\begin{figure}[b]
\rotatebox{270}{\includegraphics[height=8.5 cm]{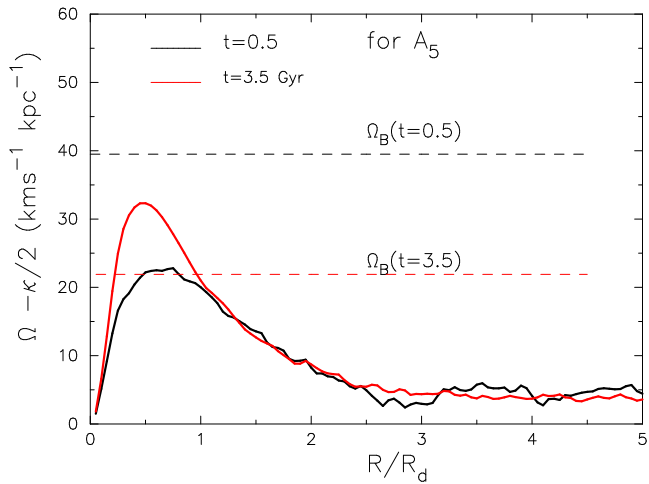}}
\caption{ILR curve plotted as function of radius for $A_5$. Two horizontal lines
mark the bar pattern speed ($\Omega_B$) at $t=0.5$ and $3.5$~Gyr.}
\label{fig:ilr}
\end{figure}

\section{Formation of stellar bars}
\label{sec:barform}

The stellar disks in all of the models are cold with $Q$ varying from $\sim 0.95 - 1.4$ 
to begin with; see table~\ref{tab:paratab}. The radial profiles of $Q(R)$ are shown in
the upper panels of Fig.~\ref{fig:QA2vstime} for type~A and type~B models. Being in 
the cold regime, the stellar disks in all models become gravitationally unstable and form 
wide-open spiral arms within a few rotation times. 
The lower panels of Fig.~\ref{fig:QA2vstime} show the time evolution of the peak of the 
$m=2$ Fourier component ($A_2$) normalized by the $m=0$ Fourier component ($A_0$). 
The value of $A_2/A_0$ sharply rises to $\sim 0.6$ within $\sim 0.5$~Gyr for all models irrespective
of the ClBs they host. As long as the stellar disk is cold, the morphological evolution
in the initial stage is nearly the same and is determined by the initial Toomre Q. The spiral arms 
grow stronger in the first $\sim 0.5$~Gyr, at which time their amplitude growth saturates and 
fragment into clumps due to non-linear effects. These spiral arms also contribute to disk heating
while dissolving \citep{Sahaetal2010}. Stellar clumps that result from the fragmented spirals migrate to
the central region and contribute to the growth of the pre-existing ClB. This can be 
visualized through the sequence of images presented in Fig.~\ref{fig:faceon} for 
type~A models. At about $0.5$~Gyr or before, all five type~A models have similar 
morphology, as determined primarily by the disk $Q$ value. 

Small differences in the initial growth rate of the $m=2$ component can be attributed to small differences 
in the initial $Q$-profile (see Fig.~\ref{fig:QA2vstime}). For example, models $A_1$
and $A_5$ have $Q=1.2$ and $Q=1.4$, respectively, at $2.5 R_d$. The stellar disk of $A_5$ is therefore
slightly warmer, and it grows the $m=2$ component more slowly than the $A_1$ disk.
However, once the initial phase is over, $A_5$ grows a stronger bar than $A_1$. 
The same dependence on $Q$ holds for the type~B models: during the initial growth phase, low-$Q$ models
tend to grow the $m=2$~component faster than their relatively high-$Q$ counterparts.
The models with very similar initial condition also tend to evolve similarly, e.g., 
$B_3$ and $B_4$, compare table~\ref{tab:paratab} and Fig.~\ref{fig:QA2vstime}.
In the final phase (after about $3.$~Gyr), the models show wide variations in the
strength of $A_2/A_0$. {\it Fig.~\ref{fig:faceon} shows the morphology of all 15 models at the
end of $\sim 3.5$~Gyr - some are clearly barred and some not.}
Slight initial variations in $Q$ values alone seem to be difficult in
providing an explanation for such wide variation in $A_2/A_0$ in the final phase. Other physical
process and/or initial parameters must be involved, as discussed in the following sections.

The models that eventually grew a bar at about $1$~Gyr (see Fig.~\ref{fig:faceon}), had
their amplitudes remain roughly constant for the next several rotation
times (see Fig.~\ref{fig:QA2vstime}). Normally, one would expect a bar to grow in amplitude
via continuously transferring angular momentum to the surrounding dark matter halo
\citep{TremaineWeinberg1984, Weinberg1985, HernquistWeinberg1992,
DebattistaSellwood2000, Athanassoula2002, SellwoodDebattista2006, Dubinskietal2009,
SahaNaab2013}. Note, the inner regions of these models are dominated by the disk and
bulge, with dark matter contributing little to the inner rotation curves; basically,
these models are maximum-disk models. The transfer of angular momentum depends
primarily on the degree of non-axisymmetry (here, the bar strength) and whether there
are adequate halo and disk particles around \citep{WeinbergKatz2007a,WeinbergKatz2007b} to 
take away the angular momentum. A part of this angular momentum from the inner region 
can be transferred to the outer disk and another part of it to the halo and bulge. 
We calculated the radial angular momentum profiles at different times and found that 
when the inner disk loses, a fraction of it goes to the outer parts and the dominant component
goes to the dark halo.

Fig.~\ref{fig:AM} illustrates the total angular momentum exchange between the full disk, bulge, 
and dark matter halo in models $A_1$ and $A_5$. In model $A_1$ (red curves), the dark matter halo absorbs 
angular momentum from the disk for an initial period of $\sim0.5$~Gyr and then saturates with 
$d({\Delta L_z})/dt =0$. Consequently, the disk of $A_1$ did not grow a bar. 
On the other hand, in model $A_5$ (blue curves), the dark matter halo gains angular 
momentum continuously and thereby facilitates in growing the bar (see Fig.~\ref{fig:faceon}). 
Since the bulge is modelled with a distribution function $f(E)$, i.e., a function of energy 
only, it always gains a small fraction of angular momentum \citep[e.g.,][]{Sahaetal2012}.

\subsection{ILR effect}
\label{sec:ilr}
In the wave-mechanics picture, a stellar bar can be thought of as a standing wave mode
- made by the linear superposition of a set of leading and trailing waves. Such a wave
could grow via swing-amplification as proposed by \cite{Toomre1981} for density waves.
The amplification of the waves depends on the corotation resonance (CR) - which plays
an important role in galactic dynamics. A CR essentially divides the whole galaxy into
two dynamically distinct parts - the region inside CR having negative energy and angular
momentum density in the wave and that outside having positive energy and angular 
momentum in the wave. So if a
wave mode loses energy and angular momentum inside CR, it will grow. An wave
incident on the CR will be partially reflected and a part will be transmitted which
will carry positive energy (if coming from the inward direction) - then for the
conservation of energy, the reflected wave from CR will be with higher (more negative)
energy (or amplitude).  This inward travelling (trailing) waves can be reflected in the
center if there no strong inner Lindblad resonance (ILR) to absorb the waves
\citep{LBK1972} - acting as a negative feedback. A centrally concentrated bulge
produces a strong ILR and prevents the feedback loop, hindering the bar growth
\citep{SellwoodEvans2001}. 

The strength of the ILR, defined as 
$[\Omega -\kappa/2]/\Omega(2R_d)$ \citep{SahaElmegreen2016}, increases as the ClBs 
become more and more compact and the peak of the ILR curve shifts inwards (see the upper 
panels of Fig.~\ref{fig:ILRA2vsR}). For the most compact ClB, $B_1$, the ILR peak lies within 
$\sim 0.1 R_d$. The differences in the ILR curves are entirely the result of the different 
initial bulge sizes and masses, because the initial disks and dark halo parameters are kept 
nearly same in each model - ideally suited for isolating the impact of bulges alone. In the 
lower panels of Fig.~\ref{fig:ILRA2vsR}, we show the radial variation of $A_2/A_0$. For both 
types of bulges, the peak location of $A_2/A_0$ occurs at smaller radii as the ILR peak shifts
inward. At the same time, the strength (maximum value of $A_2/A_0$) also decreases.

In Fig.~\ref{fig:ilr}, we show that during the initial phase of disk evolution, the model $A_5$ 
grows a bar such that it avoids a low pattern speed which would give it an ILR i.e., the 
early bar in $A_5$ starts with a pattern speed $\Omega_{B}> max[\Omega-\kappa/2]$. 
Later, the bar grows via losing angular momentum to the halo which in turn decreases 
the pattern speed. At $t=3.5$~Gyr, the pattern speed reduces to 
$\Omega_{B}=21.9$~km s$^{-1}$~kpc$^{-1}$ and it intersects the $\Omega - \kappa/2$ curve at 
two radii - producing two ILRs, allowing the formation of $x_2$ orbits in between the two radii. 
This holds true for other models forming a bar in our simulations. Both Fig.~\ref{fig:ILRA2vsR} 
and Fig.~\ref{fig:ilr} thus brings out two aspects of bar formation -

\begin{figure}
\begin{flushleft}
\rotatebox{0}{\includegraphics[height=9.0 cm]{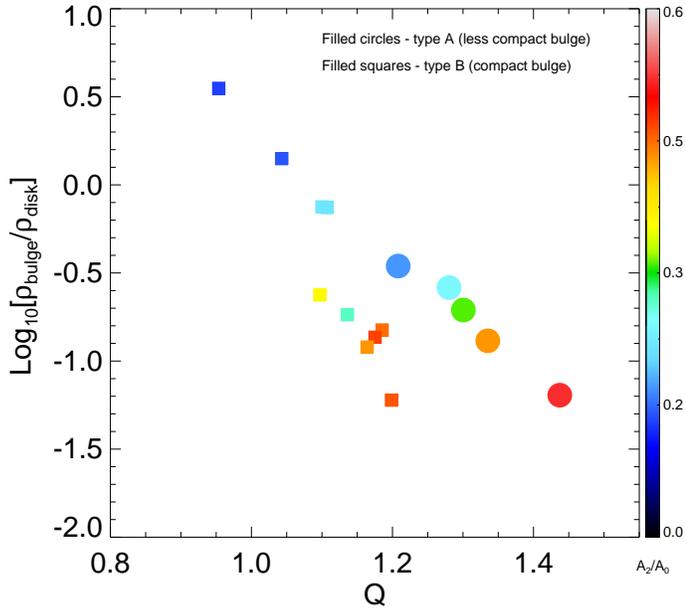}}
\caption{Bar evolution on the $Q$ - $\rho_{bulge}$ plane. The Q values are taken
at $2.5 R_d$; both $Q$ and $\rho_{bulge}/\rho_{disk}$ are computed at t=0. 
Color scale is determined by the max of $A_2/A_0$ at 3.5~Gyr. }
\label{fig:rhobulgeQ}
\end{flushleft}
\end{figure}

\begin{figure*}
\rotatebox{0}{\includegraphics[height=7.0 cm]{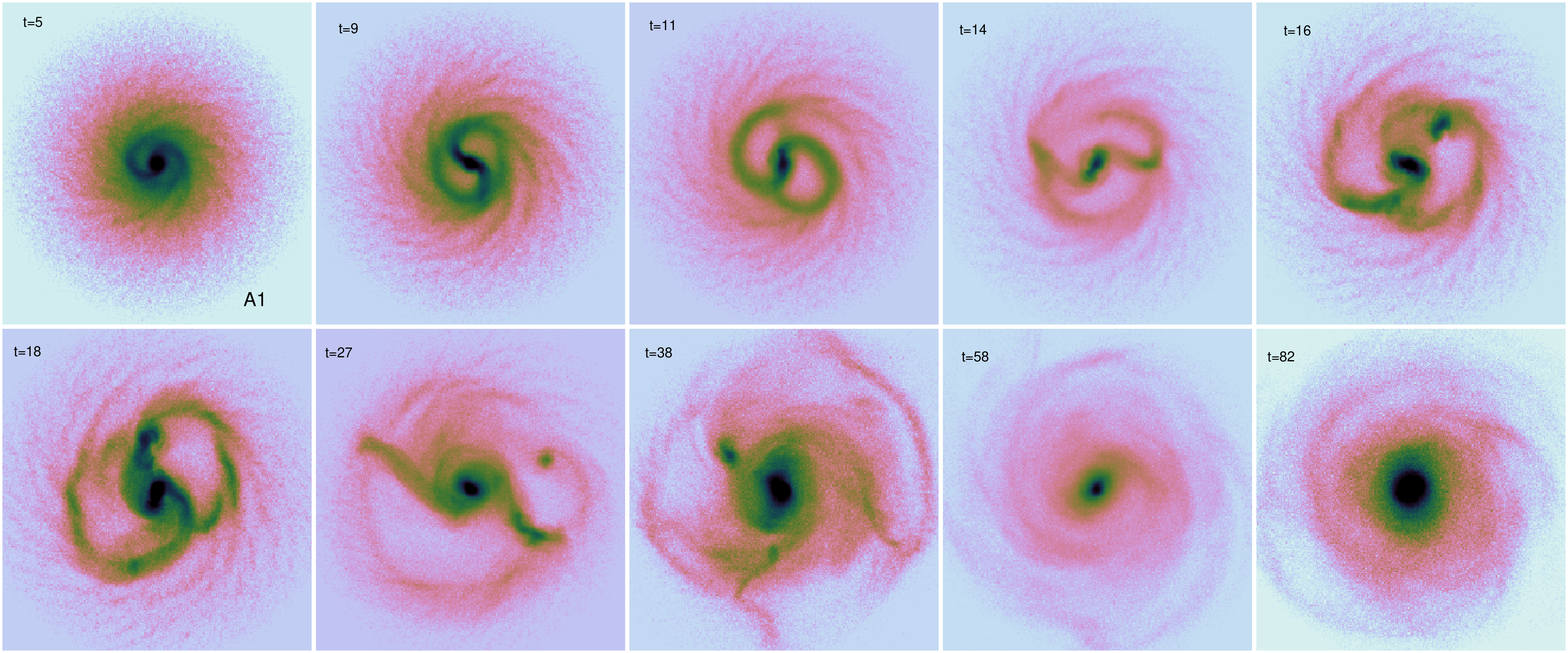}}
\caption{Density maps: close look at the first Gyr of evolution for $A_1$.}
\label{fig:faceonA5}
\end{figure*}

\noindent {\bf First:}
The final bar amplitude is found to be lower for models with higher initial ILR strength; 
for the most compact ClB, a bar is barely visible even after several rotation times.

\medskip
\noindent {\bf Second:}
The radius corresponding to the peak in the initial ILR curve is about the same as the radius of 
the maximum bar amplitude (i.e., peak of $A_2/A_0$, see Fig.~\ref{fig:ILRA2vsR}). Thus the more 
centrally concentrated bulges, which have shorter and stronger ILRs, force their incipient bars 
to be short and fast-rotating also, making them difficult to observe or short-lived.

\begin{figure*}
\rotatebox{0}{\includegraphics[height=7.0 cm]{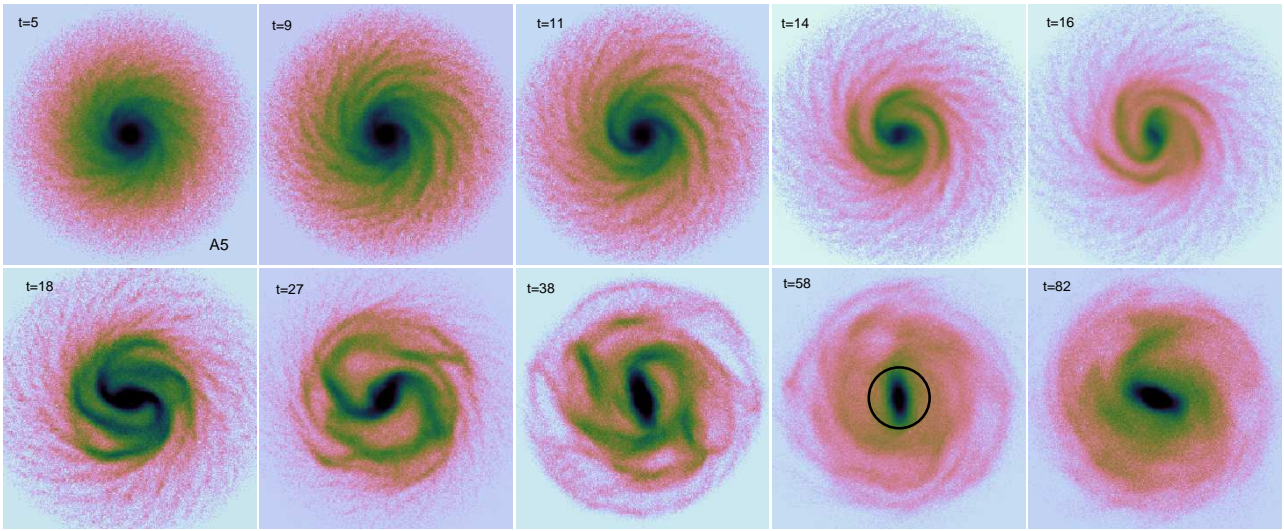}}
\caption{Density maps: close look at the first Gyr of evolution for $A_5$. The circle 
encompassing the bar at $t=58$ has a radius of $R_{bar} = 1.5 R_d$.}
\label{fig:faceonA7}
\end{figure*}

From an orbital point of view, a centrally concentrated ClB (e.g., model
$B_{1}$) might also have the effect of preventing the $x_1$ orbit families from
developing it in the first place. Any star on a highly eccentric $x_1$ orbit would pass
close to the centre and be knocked to a different orbit, which would generally be more
circular \citep{PfennigerNorman1990,Hasanetal1993}.
\noindent From both points of view, a strong ILR would be necessary to prevent a bar from
growing and we show that this is possible in the presence of a compact bulge.

\subsection{Bar strength vs bulge density and Toomre Q }
\label{sec: bulgedensity}

Fig.~\ref{fig:rhobulgeQ} summarizes how the growth of a bar depends on the
the initial Toomre Q and the normalized average bulge density.  In other words,
this plot aims at understanding the relative role of the initial Toomre Q and 
$\rho_{bulge}/\rho_{disk}$ in deciding whether a given model will evolve into 
a barred or unbarred one. Although there is no clear boundary between barred and
unbarred ones in terms of $A_2/A_0$ values, for all practical
purposes, a bar is clearly visible only when $A_2/A_0$ exceeds $\sim0.2$
\citep{SahaNaab2013}. If we consider this (i.e., $A_2/A_0 = 0.2$) as the
boundary between barred and unbarred galaxies (operational definition), an
interesting outcome arises from this figure. For type~B models, we see that only those
models with $\rho_{bulge}/\rho_{disk} > 1$ evolve into clear unbarred galaxies e.g.,
$B_{1}$, $B_{2}$. Models $B_3$ and $B_4$ are both just above the marginal case 
which  is similar to galaxies with intermediate bar types, such as SAB's with oval 
distortions (see the right panel of Fig.~\ref{fig:faceon}). Examining all our simulation 
sample, it turns out that all those model galaxies evolved to become unbarred for which 
the initial $\rho_{bulge}/\rho_{disk} > 1/\sqrt{10}$ even though they had the necessary
range of $Q$ values. In other words, a bar would preferentially form in an extended
bulge than in a compact bulge, given the same disk and halo.
Overall, there is a clear trend that as $\rho_{bulge}/\rho_{disk}$ increases, the bar strength
decreases although the disk may have Toomre Q favourable for the bar formation. 

\section{Dissection of models $A_1$ and $A_5$}
\label{sec:dissection}

The aim of this section is to carry out a detailed investigation on how the stellar disk of the
model $A_5$ eventually becomes unstable to bar formation while model $A_1$ remains stable.

\subsection{Early morphological evolution}
\label{sec:early}

In this section, we describe the early evolution of the star particles in models
$A_1$ and $A_5$; both belong to Type~A bulges - $A_1$ hosts a compact, dense
bulge while $A_5$ has a less compact bulge (see Tabel~\ref{tab:paratab}). In
Fig.~\ref{fig:faceonA5} and Fig.~\ref{fig:faceonA7}, we show the early morphological
evolution of these two models. Since the disk of $A_1$ is slightly colder than $A_5$
(see Fig.~\ref{fig:QA2vstime}), it develops spiral arms faster (within a rotation
time-scale) than $A_5$. The spiral arms reach their peak strength ($A_2/A_0 \sim
0.6$) within about 250 Myr and then break down due to non-linear processes,
forming large stellar clumps which migrate to the central region - eventually leaving
the disk in a state with negligible non-axisymmetric features. Whereas, model $A_5$
being slightly warmer, grows spiral arms rather slowly and reaches its peak value
($A_2/A_0 \sim 0.6$) around 500 Myr. Beyond this point of time, the spiral arms do not
sustain, but they decay due to the radial heating they produce in the disk. So the basic 
differences in the early evolutionary phases of models $A_1$ and $A_5$ are as follows.
In model $A_1$, spiral arms grow quickly and fragment into pieces because the initial
disk is slightly colder, forming stellar clumps and dissolving. Whereas, 
in model $A_5$, the spiral arms grow more slowly and eventually decay due to 
the slow radial heating they produce through scattering.
In the section below, we use
these observations to connect with detailed physical processes involved.

\begin{figure}[t]
\begin{flushleft}
\rotatebox{0}{\includegraphics[height=12.0 cm]{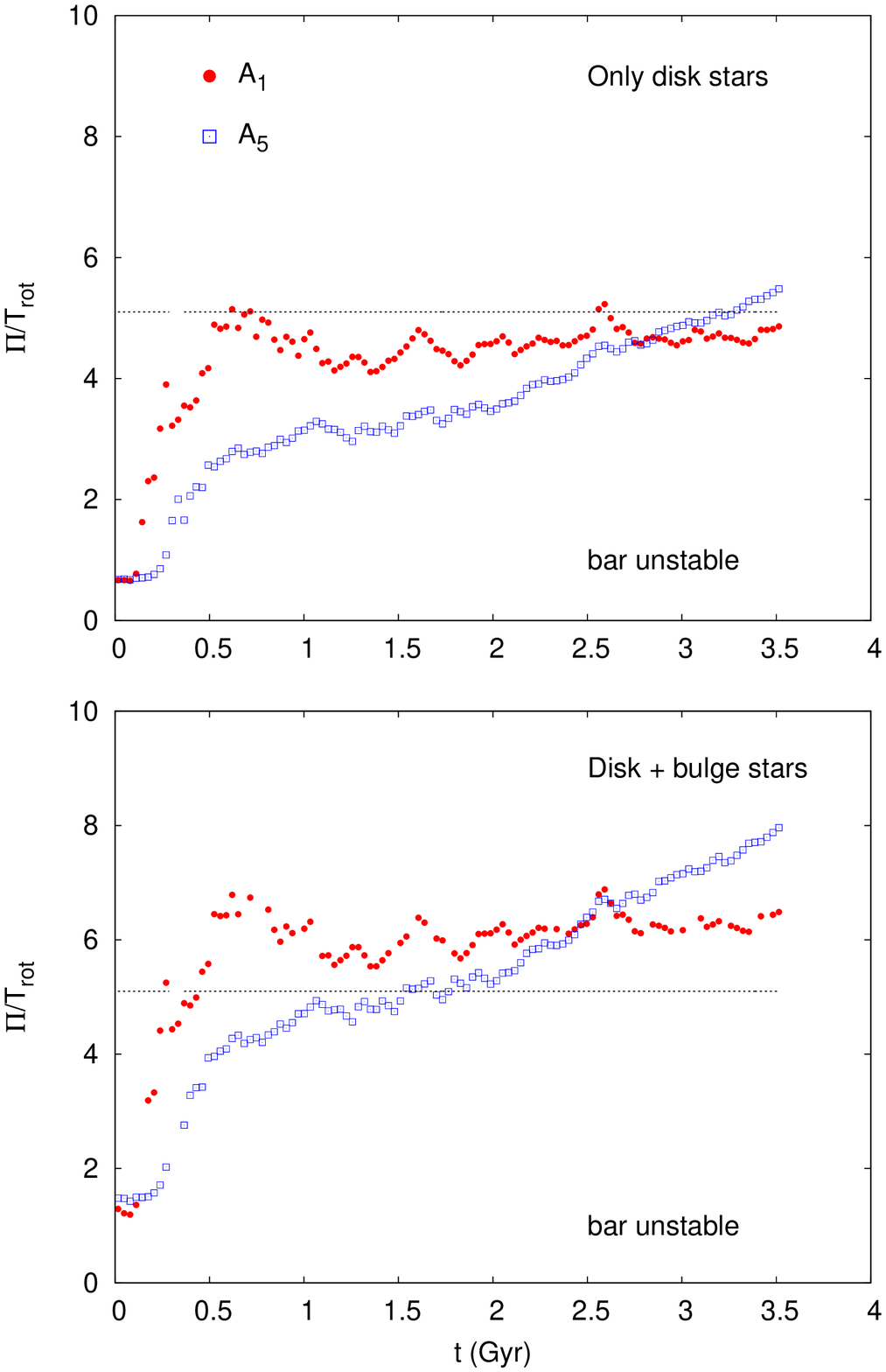}}
\caption{Ostriker-Peebles criteria for the stellar disks of models $A_1$ and $A_5$.
Upper: Only disk stars are included in the calculation. Lower: Disk plus
bulge stars for the same.  Model $A_5$ forms a bar and $A_1$ does not.}
\label{fig:tOPdisk}
\end{flushleft}
\end{figure}

\begin{figure}[b]
\begin{flushleft}
\rotatebox{0}{\includegraphics[height=12.0 cm]{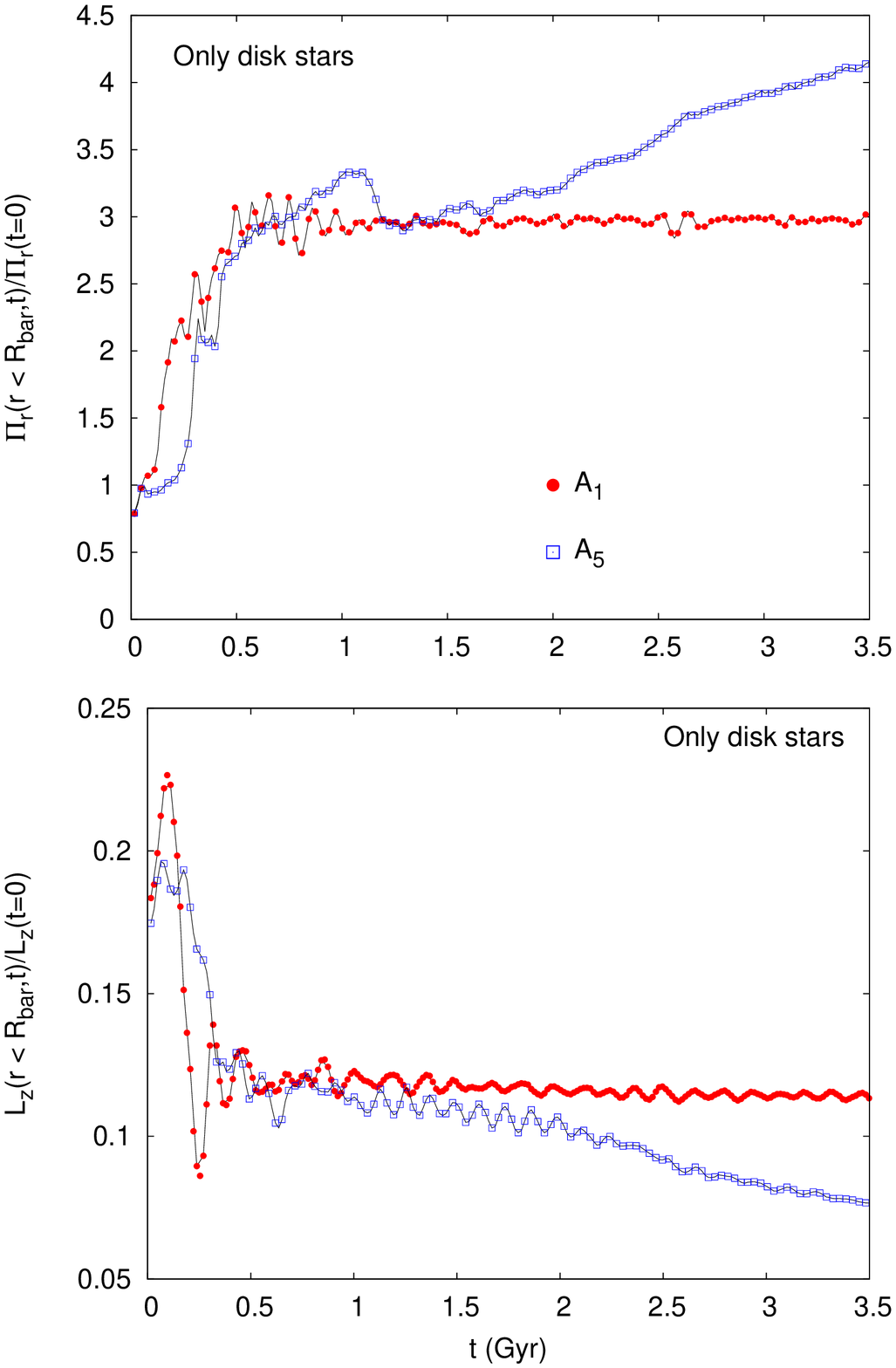}}
\caption{Upper: Time variation of the radial random kinetic energy of the stellar disk within
bar radius for models $A_1$ (remained unbarred) and $A_5$ (became barred).
Lower: Fraction of the disk total angular momentum calculated within the bar radius for both models. }
\label{fig:randErLz}
\end{flushleft}
\end{figure}

\subsection{Ostriker-Peebles criteria for bar formation}
\label{sec:OP} 

Based on the stability analysis of the Maclaurin disks \citep[see,][and references therein]{BT1987} 
and collisionless N-body simulations of galaxies, it was shown by \cite{OstrikerPeebles1973} that 
a stellar disk would go bar unstable if the ratio of the rotational kinetic energy $T_{rot}$ to potential 
energy $W$ i.e., $T_{rot}/|W|$ exceeds a critical value of $0.14 \pm 0.003$. According to \cite{Lake1983},
not only the disk, stellar bulges with $V/\sigma \sim 0.8$ might also suffer a bar instability.
We use the Ostriker-Peebles criteria to understand whether a model from our simulation sample evolves 
to become barred galaxy.
Assuming the virial theorem holds for successive snapshots in an N-body simulation, the above 
mentioned criterion can be translated in terms of the rotational and random kinetic energy alone. 
From each snapshot, we compute the kinetic energy associated with the random and mean 
motion as follows:

The random kinetic energy is given by

\begin{equation}
\Pi (r) = \Pi_{R}(r) + \Pi_{\varphi}(r) + \Pi_{z}(r),
\end{equation}

\noindent where each component of the random kinetic energy is calculated as, following \cite{BT1987}:

\begin{equation}
\Pi_{j}(r) = \sum_{i=1}^{N_r}{ m(i) \sigma_{j}(i)^{2}}
\label{randKE}
\end{equation}

In the above equation, $j = R, \varphi, z$;  $\sigma_{j}$ is the $j^{th}$ component of
the velocity dispersion, and $N_{r}$ is the number of particles, each with mass $m$ 
in a given circular annuli, $r, r+dr$ with $dr$ fixed throughout the calculation. 
We have also calculated the mean kinetic energy of the system in the given annuli.

\begin{equation}
T_{j}(r) = \sum_{i=1}^{N_r}{\frac{1}{2} m(i) {<v_{j}>}^{2}},
\label{meanKE}
\end{equation}

\noindent where $<v_{j}>$ is the $j^{th}$ component of the mean velocity of a particle.
Similar to equation above, we calculate the total mean kinetic energy associated with 
the annuli:

\begin{equation}
T_{mean}(r) = T_{R}(r) + T_{\varphi}(r) + T_{z}(r).
\end{equation}

\noindent We have verified that the total mean kinetic energy is
essentially the rotational kinetic energy. As the galaxy evolves, there are local 
variation or migrations which tend to contribute to the radial and vertical
kinetic energy - but the galaxy as a whole has no mean motion either in the radial or
vertical direction. This allows one to essentially replace $T_{mean}$ by
$T_{\varphi}=T_{rot}$. Then following the tensor-virial theorem in a steady state
\citep{BT1987}, we can write (using the trace of the kinetic and potential energy
tensors)

\begin{equation}
\Pi + 2 T_{rot} + W = 0.
\end{equation}

Then using the above equation, the Ostriker-Peebles criterion for bar instability
becomes,

\begin{equation}
\frac{T_{rot}}{|\Pi + 2 T_{rot}|} > 0.14,
\end{equation}

\noindent  or

\begin{equation}
\Pi/T_{rot} + 2 < 7.14 => \Pi/T_{rot} < 5.14
\end{equation}

In other words, if $\Pi/T_{rot} < 5.14$, then a stellar disk becomes
unstable to bar formation; $\Pi$ denotes the random kinetic energy of the stars and
$T_{rot}$ is the kinetic energy associated with the rotational motion. We compute these
quantities for each of the snapshots in our simulation using the above equations.

In Fig.~\ref{fig:tOPdisk}, we show the time evolution of $\Pi/T_{rot}$ for the two
cases: stellar disk alone and stellar disk plus bulge. If the bulge stars are excluded
from the computation, both of the cold stellar disks  would have formed a 
bar, as also suggested by the initial value of $\Pi/T_{rot} \sim 0.37$, 
 see the upper panel of  Fig.~\ref{fig:tOPdisk}.
By adding the bulges to our calculation,  the $\Pi/T_{rot}(t=0)$ increases to $\sim 1.3$ 
and $\sim 1.7$ for $A_1$ and $A_5$ respectively. Although these values have increased by a 
factor of $4 - 5$, they are still less than 5.14 - ensuring that both models qualify for 
bar-instability according to Ostriker-Peebles criteria.
But when we follow the subsequent evolution of these two models, only $A_5$ makes a bar at
the end. Clearly, the final fate of a galaxy model is not entirely decided by the initial value of
$\Pi/T_{rot} $.

\subsection{Energy and Angular momentum budget}
\label{sec:ELz}

Here, we investigate the energy and angular momentum budget in the central $1.5 R_d$
region (which encompass the bar that grows in $A_5$) of both  of these galaxies in
detail. For each of the annuli/rings of fixed size (as mentioned above), we
calculate the angular momentum as

\begin{equation}
L_z (r) = \sum_{i=1}^{N_r}{ m(i) \times [x(i) v_{y}(i)-y(i) v_x(i)]},
\end{equation}

Following Eq.\ref{randKE}, we compute the random component of the kinetic energy
associated with radial motions within the bar radius, $R_{bar} = 1.5 R_d$,
and this is repeated for every snapshot for both models. The upper panel of
Fig.~\ref{fig:randErLz} shows the time variation of radial kinetic energy ($\Pi_{r}$)
of the disk stars and the right panel shows their corresponding angular
momentum. In $A_1$, $\Pi_{r}$ increases initially more rapidly than in $A_5$. Such a
rapid increase in heating is caused by the coalescence of two giant clump-like
structures formed as a result of non-linear fragmentation of the spiral arms in the
central region, at around $t=0.25$~Gyr (see Fig.~\ref{fig:faceonA5}). Note that at this
time, the radial kinetic energy in $A_1$ has increased roughly by a factor of $2$
compared to that in $A_5$. The coalescence of the two clumps in the central
region of $A_1$ results in a sudden decrease in the angular momentum (right panel of
Fig.~\ref{fig:randErLz}). However, soon after, its angular momentum settles down
to a fixed fraction ($\sim 12 \%$) of the disk's initial angular momentum. During the
subsequent phase of evolution, model $A_1$ does not undergo any major change; both
radial kinetic energy and z-component of angular momentum stay nearly constant.
Whereas in model $A_5$, the inner stellar disk steadily loses angular momentum as
would be expected for a disk that is growing a bar. The bar
facilitates a steady loss of angular momentum from the inner disk accompanied by a
steady increase in the radial kinetic energy (blue points in Fig.~\ref{fig:randErLz}).

To summarise, in an isolated galaxy under virial equilibrium, a rapid (or
non-adiabatic) loss of angular momentum is also accompanied by a non-adiabatic heating
of the inner disk. As a result, orbital structure in the inner region has very little
time to respond to such rapid non-adiabatic change in the disk angular momentum. This
is probably causing the prevention of a bar in model $A_1$. In other words, {\it the key to 
``not forming a bar'' is to find a way such that the central part of the disk undergoes a 
sudden loss of angular momentum associated with simultaneous heating to a high degree.}
This can be achieved in a galaxy model having a cold stellar disk and a compact bulge whose 
average density is greater than or comparable to the disk density within the bulge half-mass radius.  

\section{Discussion and Conclusions}
\label{sec:discussion}
The models $A_1$, $B_1$ and $B_2$ have the most compact ClBs and they all evolve into galaxies
with final $A_2/A_0 < 0.2$. Stronger $m=2$ modes grow in models with more extended
bulges of the same mass, i.e., lower density bulges. The orbits of the bulge stars
are themselves hot thermal orbits and not circular, and that what is important for bar 
formation in the disk is the cold orbits of the subcomponent of the "bulge-region" stars 
that is in the disk. Additionally, a hotter bulge is not easily deformed by perturbations
in the disk, it is non-reactive and therefore disk perturbations can not amplify very much
and grow into a disk bar.    
The bar needs $x_1$ orbits to reinforce it, and
when the bar is initially weak, its pattern speed cannot be lower than the peak in
$\Omega-\kappa/2$ because then $x_2$ orbits would form between the two ILR radii. Thus
seed bars may come and go repeatedly at high angular frequency and small radius when
the ILR peak is large, but such tiny bars are not typically classified as barred
galaxies.

Our simulations are pure collisionless in nature, i.e., without any dissipative
component such as cold gas. The presence of a gas component in N-body simulations of
disk galaxies is known to contribute to the weakening of the
already-present bar \citep{Berentzenetal1998,Athanassoulaetal2013}. Added to this are
the central mass concentrations (CMC) and super-massive black holes (SMBH) at the
galactic centre which have a destructive effect on the galactic bar
\citep{Hasanetal1993,Bournaudetal2005,HozumiHernquist2005}. Nonetheless as shown by
\cite{Athanassoulaetal2005}, a bar is hard to destroy completely either by CMCs or
SMBHs.
\smallskip

The main results from our work are as follows:

{\noindent \bf 1.} Our simulations, in essence, show how a cold stellar disk that is prone to bar
instability prevents a bar from forming in the presence of a compact and highly dense 
classical bulge.
\smallskip

{\noindent  \bf 2.}
Based on pure stellar dynamical effects, we suggest that the recipe to
prevent frequent bar formation in simulations is to let the central few kpc region of
the stellar disk undergo a non-adiabatic (rapid) loss of angular momentum accompanied by
a simultaneous rise in the radial kinetic energy.
\smallskip

{\noindent  \bf 3.}
The analyses from our simulations suggest that model galaxies that prevent bar 
formation and remained completely unbarred at later stages of evolution, had their 
initial bulge densities greater than or comparable to the disk density measured
within the bulge half-mass radii. 

\smallskip
{\noindent  \bf 4.}
The bars that formed in our simulations during the early phase of the evolution had
their pattern speed always greater than the maximum of $\Omega-\kappa/2$, i.e., the
early bars avoided the ILR. The barred galaxies also had a low ratio of random
energy to rotational energy, less than $\sim 5.14$, for over a Gyr initially while the
non-barred galaxy models reached a high ratio fairly early, in less than half a Gyr.  
The prolonged period of relatively low random energy allowed the bar to form over several
rotation periods.

\section*{acknowledgement} 
\noindent The numerical simulations were carried out at the IUCAA HPC cluster. The authors thank the
anonymous referee for useful comments.


\begin{thebibliography}{57}
\expandafter\ifx\csname natexlab\endcsname\relax\def\natexlab#1{#1}\fi

\bibitem[{{Athanassoula}(2002)}]{Athanassoula2002}
{Athanassoula}, E. 2002, \apjl, 569, L83

\bibitem[{{Athanassoula}(2003)}]{Athanassoula2003}
---. 2003, \mnras, 341, 1179

\bibitem[{{Athanassoula} {et~al.}(2005){Athanassoula}, {Lambert}, \&
  {Dehnen}}]{Athanassoulaetal2005}
{Athanassoula}, E., {Lambert}, J.~C., \& {Dehnen}, W. 2005, \mnras, 363, 496

\bibitem[{{Athanassoula} {et~al.}(2013){Athanassoula}, {Machado}, \&
  {Rodionov}}]{Athanassoulaetal2013}
{Athanassoula}, E., {Machado}, R.~E.~G., \& {Rodionov}, S.~A. 2013, \mnras,
  429, 1949

\bibitem[{{Athanassoula} \& {Misiriotis}(2002)}]{Athanamisi2002}
{Athanassoula}, E., \& {Misiriotis}, A. 2002, \mnras, 330, 35

\bibitem[{{Barazza} {et~al.}(2008){Barazza}, {Jogee}, \&
  {Marinova}}]{Barazzaetal2008}
{Barazza}, F.~D., {Jogee}, S., \& {Marinova}, I. 2008, \apj, 675, 1194

\bibitem[{{Barnes} \& {Hernquist}(1991)}]{BarnesHerquist1991}
{Barnes}, J.~E., \& {Hernquist}, L.~E. 1991, \apjl, 370, L65

\bibitem[{{Berentzen} {et~al.}(1998){Berentzen}, {Heller}, {Shlosman}, \&
  {Fricke}}]{Berentzenetal1998}
{Berentzen}, I., {Heller}, C.~H., {Shlosman}, I., \& {Fricke}, K.~J. 1998,
  \mnras, 300, 49

\bibitem[{{Binney} \& {Tremaine}(1987)}]{BT1987}
{Binney}, J., \& {Tremaine}, S. 1987, {Galactic dynamics}, ed. {Binney, J.~\&
  Tremaine, S.}

\bibitem[{{Bournaud} \& {Combes}(2002)}]{BournaudCombes2002}
{Bournaud}, F., \& {Combes}, F. 2002, \aap, 392, 83

\bibitem[{{Bournaud} {et~al.}(2005){Bournaud}, {Combes}, \&
  {Semelin}}]{Bournaudetal2005}
{Bournaud}, F., {Combes}, F., \& {Semelin}, B. 2005, \mnras, 364, L18

\bibitem[{{Ceverino} \& {Klypin}(2007)}]{Ceverinoklypin2007}
{Ceverino}, D., \& {Klypin}, A. 2007, \mnras, 379, 1155

\bibitem[{{Combes} \& {Sanders}(1981)}]{CombesSanders1981}
{Combes}, F., \& {Sanders}, R.~H. 1981, \aap, 96, 164

\bibitem[{{Debattista} \& {Sellwood}(1998)}]{DebattistaSellwood1998}
{Debattista}, V.~P., \& {Sellwood}, J.~A. 1998, \apjl, 493, L5

\bibitem[{{Debattista} \& {Sellwood}(2000)}]{DebattistaSellwood2000}
---. 2000, \apj, 543, 704

\bibitem[{{Dubinski} {et~al.}(2009){Dubinski}, {Berentzen}, \&
  {Shlosman}}]{Dubinskietal2009}
{Dubinski}, J., {Berentzen}, I., \& {Shlosman}, I. 2009, \apj, 697, 293

\bibitem[{{Elmegreen} {et~al.}(1991){Elmegreen}, {Sundin}, {Sundelius}, \&
  {Elmegreen}}]{Elmegreen1991}
{Elmegreen}, D.~M., {Sundin}, M., {Sundelius}, B., \& {Elmegreen}, B. 1991,
  \aap, 244, 52

\bibitem[{{Eskridge} {et~al.}(2000){Eskridge}, {Frogel}, {Pogge}, {Quillen},
  {Davies}, {DePoy}, {Houdashelt}, {Kuchinski}, {Ram{\'{\i}}rez}, {Sellgren},
  {Terndrup}, \& {Tiede}}]{Eskridgeetal2000}
{Eskridge}, P.~B., {Frogel}, J.~A., {Pogge}, R.~W., {et~al.} 2000, \aj, 119,
  536

\bibitem[{{Evans}(1993)}]{Evans93}
{Evans}, N.~W. 1993, \mnras, 260, 191

\bibitem[{{Gerin} {et~al.}(1990){Gerin}, {Combes}, \&
  {Athanassoula}}]{Gerin1990}
{Gerin}, M., {Combes}, F., \& {Athanassoula}, E. 1990, \aap, 230, 37

\bibitem[{{Goldreich} \& {Tremaine}(1979)}]{Goldreich-Tremaine1979}
{Goldreich}, P., \& {Tremaine}, S. 1979, \apj, 233, 857

\bibitem[{{Grosb{\o}l} {et~al.}(2004){Grosb{\o}l}, {Patsis}, \&
  {Pompei}}]{Grosboletal2004}
{Grosb{\o}l}, P., {Patsis}, P.~A., \& {Pompei}, E. 2004, \aap, 423, 849

\bibitem[{{Hasan} {et~al.}(1993){Hasan}, {Pfenniger}, \&
  {Norman}}]{Hasanetal1993}
{Hasan}, H., {Pfenniger}, D., \& {Norman}, C. 1993, \apj, 409, 91

\bibitem[{{Hernquist} \& {Weinberg}(1992)}]{HernquistWeinberg1992}
{Hernquist}, L., \& {Weinberg}, M.~D. 1992, \apj, 400, 80

\bibitem[{{Hohl}(1971)}]{hohl71}
{Hohl}, F. 1971, \apj, 168, 343

\bibitem[{{Holley-Bockelmann} {et~al.}(2005){Holley-Bockelmann}, {Weinberg}, \&
  {Katz}}]{Holley-Bockelmannetal2005}
{Holley-Bockelmann}, K., {Weinberg}, M., \& {Katz}, N. 2005, \mnras, 363, 991

\bibitem[{{Hozumi} \& {Hernquist}(2005)}]{HozumiHernquist2005}
{Hozumi}, S., \& {Hernquist}, L. 2005, \pasj, 57, 719

\bibitem[{{King}(1966)}]{King1966}
{King}, I.~R. 1966, \aj, 71, 64

\bibitem[{{Kuijken} \& {Dubinski}(1995)}]{KD1995}
{Kuijken}, K., \& {Dubinski}, J. 1995, \mnras, 277, 1341

\bibitem[{{Lake}(1983)}]{Lake1983}
{Lake}, G. 1983, \apj, 264, 408

\bibitem[{{Lynden-Bell} \& {Kalnajs}(1972)}]{LBK1972}
{Lynden-Bell}, D., \& {Kalnajs}, A.~J. 1972, \mnras, 157, 1

\bibitem[{{McMillan} \& {Dehnen}(2007)}]{McMillan2007}
{McMillan}, P.~J., \& {Dehnen}, W. 2007, \mnras, 378, 541

\bibitem[{{Men{\'e}ndez-Delmestre} {et~al.}(2007){Men{\'e}ndez-Delmestre},
  {Sheth}, {Schinnerer}, {Jarrett}, \& {Scoville}}]{MenendezDelmestreetal2007}
{Men{\'e}ndez-Delmestre}, K., {Sheth}, K., {Schinnerer}, E., {Jarrett}, T.~H.,
  \& {Scoville}, N.~Z. 2007, \apj, 657, 790

\bibitem[{{Miller} {et~al.}(1970){Miller}, {Prendergast}, \&
  {Quirk}}]{miller70}
{Miller}, R.~H., {Prendergast}, K.~H., \& {Quirk}, W.~J. 1970, \apj, 161, 903

\bibitem[{{Miwa} \& {Noguchi}(1998)}]{MiwaNoguchi1998}
{Miwa}, T., \& {Noguchi}, M. 1998, \apj, 499, 149

\bibitem[{{Noguchi}(1987)}]{Noguchi1987}
{Noguchi}, M. 1987, \mnras, 228, 635

\bibitem[{{Ostriker} \& {Peebles}(1973)}]{OstrikerPeebles1973}
{Ostriker}, J.~P., \& {Peebles}, P.~J.~E. 1973, \apj, 186, 467

\bibitem[{{Pfenniger} \& {Norman}(1990)}]{PfennigerNorman1990}
{Pfenniger}, D., \& {Norman}, C. 1990, \apj, 363, 391

\bibitem[{{Polyachenko}(2013)}]{Polyachenko2013}
{Polyachenko}, E.~V. 2013, Astronomy Letters, 39, 72

\bibitem[{{Romano-D{\'{\i}}az} {et~al.}(2008){Romano-D{\'{\i}}az}, {Shlosman},
  {Heller}, \& {Hoffman}}]{Romano-Diazetal2008}
{Romano-D{\'{\i}}az}, E., {Shlosman}, I., {Heller}, C., \& {Hoffman}, Y. 2008,
  \apjl, 687, L13

\bibitem[{{Saha} \& {Elmegreen}(2016)}]{SahaElmegreen2016}
{Saha}, K., \& {Elmegreen}, B. 2016, \apjl, 826, L21

\bibitem[{{Saha} {et~al.}(2012){Saha}, {Martinez-Valpuesta}, \&
  {Gerhard}}]{Sahaetal2012}
{Saha}, K., {Martinez-Valpuesta}, I., \& {Gerhard}, O. 2012, \mnras, 421, 333

\bibitem[{{Saha} \& {Naab}(2013)}]{SahaNaab2013}
{Saha}, K., \& {Naab}, T. 2013, \mnras, 434, 1287

\bibitem[{{Saha} {et~al.}(2010){Saha}, {Tseng}, \& {Taam}}]{Sahaetal2010}
{Saha}, K., {Tseng}, Y., \& {Taam}, R.~E. 2010, \apj, 721, 1878

\bibitem[{{Sellwood} \& {Debattista}(2006)}]{SellwoodDebattista2006}
{Sellwood}, J.~A., \& {Debattista}, V.~P. 2006, \apj, 639, 868

\bibitem[{{Sellwood} \& {Evans}(2001)}]{SellwoodEvans2001}
{Sellwood}, J.~A., \& {Evans}, N.~W. 2001, \apj, 546, 176

\bibitem[{{Sellwood} \& {Wilkinson}(1993)}]{SellwoodWilkinson1993}
{Sellwood}, J.~A., \& {Wilkinson}, A. 1993, Reports on Progress in Physics, 56,
  173

\bibitem[{{Shen} \& {Sellwood}(2004)}]{ShenSellwood2004}
{Shen}, J., \& {Sellwood}, J.~A. 2004, \apj, 604, 614

\bibitem[{{Sheth} {et~al.}(2012){Sheth}, {Melbourne}, {Elmegreen}, {Elmegreen},
  {Athanassoula}, {Abraham}, \& {Weiner}}]{sheth12}
{Sheth}, K., {Melbourne}, J., {Elmegreen}, D.~M., {et~al.} 2012, \apj, 758, 136

\bibitem[{{Sheth} {et~al.}(2008){Sheth}, {Elmegreen}, {Elmegreen}, {Capak},
  {Abraham}, {Athanassoula}, {Ellis}, {Mobasher}, {Salvato}, {Schinnerer},
  {Scoville}, {Spalsbury}, {Strubbe}, {Carollo}, {Rich}, \& {West}}]{sheth08}
{Sheth}, K., {Elmegreen}, D.~M., {Elmegreen}, B.~G., {et~al.} 2008, \apj, 675,
  1141

\bibitem[{{Springel} {et~al.}(2001){Springel}, {Yoshida}, \&
  {White}}]{Springeletal2001}
{Springel}, V., {Yoshida}, N., \& {White}, S.~D.~M. 2001, \na, 6, 79

\bibitem[{{Toomre}(1964)}]{Toomre1964}
{Toomre}, A. 1964, \apj, 139, 1217

\bibitem[{{Toomre}(1981)}]{Toomre1981}
{Toomre}, A. 1981, in Structure and Evolution of Normal Galaxies, ed.
  {S.~M.~Fall \& D.~Lynden-Bell}, 111--136

\bibitem[{{Tremaine} \& {Weinberg}(1984)}]{TremaineWeinberg1984}
{Tremaine}, S., \& {Weinberg}, M.~D. 1984, \mnras, 209, 729

\bibitem[{{Weinberg}(1985)}]{Weinberg1985}
{Weinberg}, M.~D. 1985, \mnras, 213, 451

\bibitem[{{Weinberg} \& {Katz}(2007{\natexlab{a}})}]{WeinbergKatz2007a}
{Weinberg}, M.~D., \& {Katz}, N. 2007{\natexlab{a}}, \mnras, 375, 425

\bibitem[{{Weinberg} \& {Katz}(2007{\natexlab{b}})}]{WeinbergKatz2007b}
---. 2007{\natexlab{b}}, \mnras, 375, 460

\end{thebibliography}


\end{document}